# Context Ontology Implementation for Smart Home

Tam Van Nguyen, Wontaek Lim, Huy Anh Nguyen, Deokjai Choi and Chilwoo Lee
*Department Of Computer Engineering, Chonnam National University*
*300 Yongbong-dong, Buk-gu, Gwangju, 500-757, Korea*
*Email: vantam@gmail.com, rltk139@daum.net, anhhuy@gmail.com, dchoi@chonnam.ac.kr*

## Abstract

*Context awareness is one of the important fields in ubiquitous computing. Smart Home, a specific instance of ubiquitous computing, provides every family with opportunities to enjoy the power of hi-tech home living. Discovering that relationship among user, activity and context data in home environment is semantic, therefore, we apply ontology to model these relationships and then reason them as the semantic information. In this paper, we present the realization of smart home's context-aware system based on ontology. We discuss the current challenges in realizing the ontology context base. These challenges can be listed as collecting context information from heterogeneous sources, such as devices, agents, sensors into ontology, ontology management, ontology querying, and the issue related to environment database explosion.*

## 1. Introduction

Smart home is getting popular nowadays. To realize the human-centric technology in a real home, we need a ubiquitous computing infrastructure which can be aware of contexts and provide appropriate data and services. The society that we are living today is focusing on the direction of a human-oriented society. The ubiquitous computing infrastructure generally consists of sensor, middleware and application layers. Each layer is structurally separated and run independently. The sensor layer obtains primitive information from environment monitoring using various sensors and manual input by users. The middleware layer stores and analyzes primitive information and make a decision to current contexts. The application layer provides users with related data and services according to current contexts.

A key to realizing this context aware system is the use of a set of common ontologies that support the communication and representation. Ontology is a common knowledge that represents all information of the ubiquitous computing environment [1, 2, 3, 4]. It is sharable between the middleware and applications so that application developers can make rules to define application context without being dependent on a specific context model of middleware. To adapt dynamically to changing contexts, applications should be able to receive dynamic information according to context changes.

There are some issues in applying ontology for context aware systems. The first issue is the definition of home ontology in a specific environment, which is not clear up to now. The second one is the difficulty in ontology storage. The last but not the least challenge is the use of ontology in querying to satisfy the need of users in smart home. Our paper can help construct the ontology for smart home infrastructure in which interacts among sensor, middleware, and application layer. Application developers can build applications easily by using our ontology, so it is expected to advance the realization of the human-centric home ubiquitous computing technology.

The rest of this paper is organized as follows. First, we introduce the brief survey of recently related works in section 2. After that, the context ontology implementation process part and the implementation part are described in Section 3, Section 4, respectively. Finally we give our conclusion and discuss the future works in Section 5.

## 2. Related works

In this section, we briefly describe the little survey of recently works related to our research also using ontology. Our work is a part of UTOPIA, our on-going project, where we realize the ontology-based context aware. UTOPIA (UbiquiTous cOmPutIng Architecture), is the framework for supporting the application context [5]. UTOPIA allows each application to define middleware-independent contexts using shared ontology so that it can help developers create applications easily.

These systems developed in the past aim to support pervasive computing, such as Cooltown [2] and Context Toolkit [3], having made progress in various aspects of pervasive computing, they offer only weak support for knowledge sharing and context reasoning. A significant source of this weakness is that they are not built on a foundation of common ontologies with explicit semantic representation. However, none have taken advantage of the semantics of spatial relations in reasoning about context (i.e., information that describes the whole physical space that surrounds a particular location and its relationship to other locations).

In SOCAM [1], application developers should create pre-defined rules in a file for context decision and pre-load them into context reasoning module in middleware. These steps can not be done without administrator's operation. Some infrastructures can not provide both context trigger and context query while they provide one of the two functions independent on middleware. The Context Broker also supports context query function based on its own inference rules but does not support context decision function.

Home environment has its own characteristics. Therefore, there is a need to create the specific ontology for specific space, smart home. Furthermore, some previous systems often implemented context as simple programming language objects (e.g., Java class objects) or informally described in documentation. Because these representations require the establishment of a priori low-level implementation agreement among programs that wish to share information, they cannot facilitate knowledge sharing in an open and dynamic environment. In order to facilitate the sharing of contextual knowledge, we believe ontologies of context related information must be defined in order to provide a set of common vocabularies with shared semantics.

## 3. Context ontology implementation process

In this section, we present the context ontology implementation process. We then introduce RDF and OWL as the background knowledge to realize the ontology. After that, we introduce the context reality, ontology meta-data in turn.

### 3.1. Context ontology implementation overview

In the artificial intelligence literature, ontology is a formal, explicit description of concepts in a particular domain of discourse. It provides a vocabulary for representing domain knowledge and for describing specific situations in a domain.

Context can be defined as the semantic representation of real-world knowledge in a machine understandable format. An ontology-based approach for context modeling lets us describe contexts semantically and share common understanding of the structure of contexts among users, devices, and services. The context implementation process is briefly introduced in Figure 1. In smart home reality, we extract and model the necessary information as hierarchical tree. Then, we realize the model in meta model format as RDF or OWL. After that, the realization will be integrated into context aware system; the part has close relationship with Home Appliance Control System, which plays the main role in Smart Home Automation.

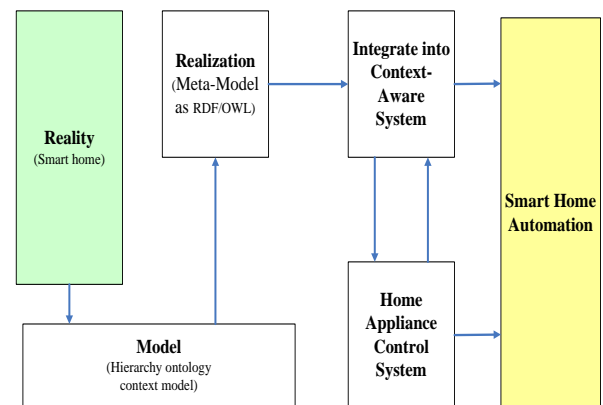

**Figure 1. The context ontology implementation process**

### 3.2. RDF and OWL overview

RDF (Resource Description Framework) is a language for representing meta-data [6]. A RDF statement is typically expressed as" resource - property - value" - triple, commonly written as P(R, V): A resource R has a property P with value V. These triples can also be seen as object-attribute-value triple. Statements can also be ex-pressed as graphs with node for resources and values where directed edges represent the properties.

OWL (Web Ontology Language) is a language for defining ontologies [7]. Ontology is referred as the shared understanding of some domains, which is often conceived as a set of entities, relations, functions, axioms and instances. The OWL language builds on the DAML+OIL language [8] and both are layered on top of on the standard RDF/RDFS triple data model as mentioned in the RDF introduction [6]. OWL has many advantages. First, it is much expressive compared to other ontology languages such as RDFS [9]. Second, it has the capability of supporting semantic interoperability to exchange and share context knowledge between different systems, i.e., contexts can be exchanged and understood between different systems in various domains; and enabling automated reasoning to be used by automated processes. Moreover, DAML+OIL are integrated into OWL to become an open W3C standard. Therefore, we choose OWL to realize our context model and define our context ontology.

### 3. 3. Context ontology model for Smart Home

Many researchers have attempted to define context by enumerating examples of contexts. Schilit classified context into three categories [10].Computational context, such as network connectivity, communication costs, and communication bandwidth, and nearby resources such as printers, displays, and workstations. User contexts can be the user's profile, location, and people nearby, even the current social situation. Meanwhile, physical contexts are lighting, noise levels, traffic conditions, and temperature. Some other researchers try to formally define context. Schmidt et al. define context as "knowledge about the user's and IT device's state, including surroundings, situation, and to a less extent, location" [11]. Dey defines context as "any information that can be used to characterize the situation of an entity. An entity can be a person, place, or object that is considered relevant to the interaction between a user and an application including the user and applications themselves" [12].

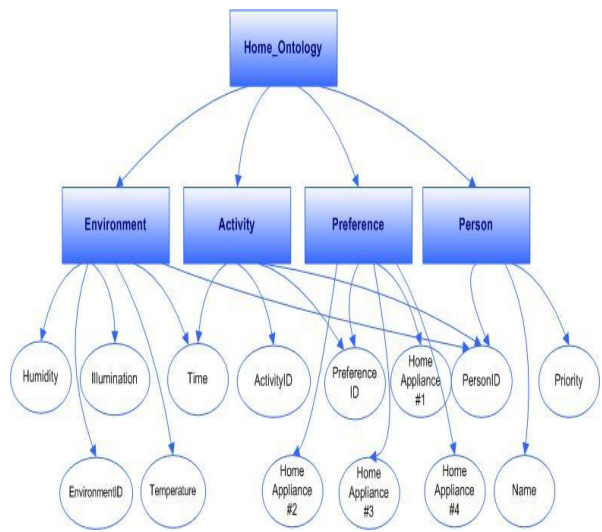

**Figure 2. Hierarchy ontology context model**

In our approach, we recognize that time is also an important and natural context for many applications. Since it is hard to fit into any of the above three kinds of context, we propose to add a more context category. Additionally, priority is the parameter used to rank the preferences of users. Therefore, time and person's priority are added as the important and natural context information for many applications. More importantly, when the computing, user and physical contexts are recorded across a time span, we obtain a context history, which could also be useful for some applications. Figure 2 shows the hierarchy ontology organization. This figure shows the ontology structure and relationship of environmental data in our context aware system. The circle shapes represent the attributes whereas the rectangle shapes stand for the ontology classes. As standard methodologies for discovery and service invocation can be implemented using agent frameworks, a standard representation of context meta-data is also necessary. Semantic ontology can provide such a standard representation.

## 4. Implementation and achievements

In this section, we present our implementation. After the implementation introduction, we show some achievement results from this.

### 4.1. Implementation

We implement both home server and web server, two important components in smart home. While home server is responsible for controlling home appliances, web server is the place where we can store ontology. The integration ontology into context aware system is installed in web server. The home server, an embedded board, sends the environmental information to web server by using JMS [13]. Currently, in the recognition part, testers wear RFID tags and the RFID reader will detect someone gets in or gets out. When converting raw sensor data into ontologies, we use DOM model to create the ontologies because RDF/OWL is from XML. Additionally, the environment sensors we use in Smart Home are TIP 700CM (Maxfor Inc.) [14]. Figure 3a shows the TIP 700CM sensor while Figure 3b shows the user interface of sensor monitoring module.

In smart home reality, the environment changes slightly. Thus, in order to prevent the population explosion of environmental data, we must exclude the duplicate data. We assume that the collection of environment information is a set whose difference can be calculated by using dissimilarity formulas.

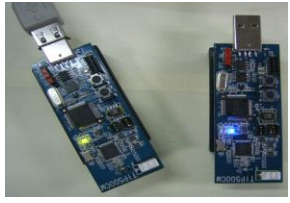

**(a)**

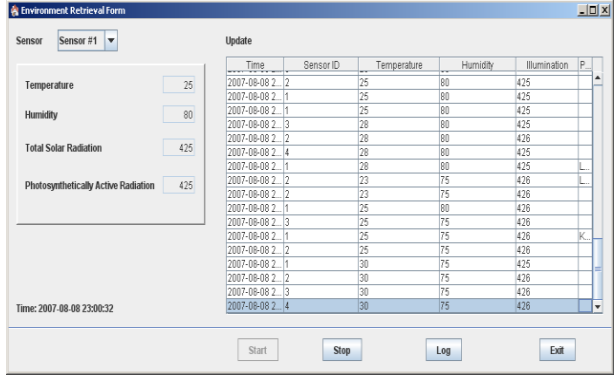

**(b)**

**Figure 3. The TIP 700CM sensor and the screen capture of monitoring environmental data**

The similarity of two environmental entities *i* and *j* is taken to be inversely related to their distance, *D(i, j)*, which is computed by

$$D(i,j) = \left[ \sum_{k=1}^{n} \left| X_{ik} - X_{jk} \right|^2 \right]^{\frac{1}{2}}$$

where *n* is the number of environmental information, whereas $X_{ik}$ and $X_{jk}$ are the values of kth element for entity i and j respectively. $X_{ik}$ is the value of current state while $X_{jk}$ is the value of nearest known state of each environmental parameter (e.g. humidity (%), temperature (°C), illumination (lux), date (year-month-day)). If a value has not changed significantly since the last time it was updated, it is not stored to the database. Significant level depends on the alteration of sensor information being monitored. The environment change as above formula changing by 1% is not significant; however, it is really a critical problem if the change is over 50%. For example, if homeowner enters his empty house and the environmental information doesn't change, as follow the formula, we will have D(i,j) = 100%. The environmental information will be stored in the case of the dissimilarity larger than the threshold. Table I shows the thresholds we infer from reality.

**TABLE I**

**THRESHOLDS FOR ENVIRONMENT**

| Factor | Temperature | Illumination | Humidity | Human presence |
|---|---|---|---|---|
| Threshold | 0.1 | 0.5 | 0.35 | in/out |

The shared ontology base provides consistent context knowledge storage. As described above, we have a context model. The shared ontology stores instances' values into a file which can be accessed by all context objects that evaluate their rules. We used the Protégé ontology editor to build context modeling and create an OWL file, and update instances' values by using API of Jena [15] and Protégé OWL reasoner [16].

### 4.2. Achievement results

The realization of ontology based on our definition of home ontology above is relatively eligible for Smart Home. The extraction of all sort of ontologies are shown in Figure 4. As seen in this figure, the ontologies are modeled in RDF/OWL language.

| **Environment** |
|---|
| <EnvironmentInfo rdf:ID="_070411115500"><br>    <Humidity rdf:datatype="xsd:double">30</Humidity><br>    <Temperature rdf:datatype="xsd:double">20</Temperature><br>    <Illumination rdf:datatype="xsd:double">400</Illumination><br>    <Date rdf:datatype="xsd:date">2007-04-11</Date>`<br>    <hasTime rdf:resource="#_115500"/><br>    <personIn rdf:resource="#Father"/><br></EnvironmentInfo> |
| **Activity** |
| <owl:distinctMembers rdf:parseType="Collection"><br>    <Activities rdf:ID="ofChildren"><br>        <When rdf:resource="#_180000"/><br>        <Who rdf:resource="#Son"/><br>        <Do rdf:resource="#Self-study"/><br>    </Activities><br>    <Activities rdf:ID="ofFather"><br>        <When rdf:resource="#_200000"/><br>        <Do rdf:resource="#Entertain"/><br>        <Who rdf:resource="#Father"/><br>    </Activities><br></owl:distinctMembers> |
| **Preference** |
| <owl:distinctMembers rdf:parseType="Collection"><br>    <Preferences rdf:ID="Entertain"><br>        <Projector rdf:datatype="xsd:boolean">true</Projector><br>        <AirConditioner rdf:datatype="xsd:boolean">true</AirConditioner><br>        <Light rdf:datatype="xsd:boolean">true</Light><br>        <TV rdf:datatype="xsd:boolean">true</TV><br>    </Preferences><br>    <Preferences rdf:ID="Self-study"><br>        <Projector rdf:datatype="xsd:boolean">true</Projector><br>        <AirConditioner rdf:datatype="xsd:boolean">true</AirConditioner><br>        <Light rdf:datatype="xsd:boolean">true</Light><br>        <TV rdf:datatype="xsd:boolean">false</TV><br>    </Preferences><br></owl:distinctMembers> |
| **Person** |
| <owl:distinctMembers rdf:parseType="Collection"><br>    <Person rdf:ID="Father"><br>        <name rdf:datatype="xsd:string">John</name><br>        <hasPriority rdf:datatype="xsd:positiveInteger">8</hasPriority><br>    </Person><br>    <Person rdf:ID="Son"><br>        <name rdf:datatype="xsd:string">Tom</name><br>        <hasPriority rdf:datatype="xsd:positiveInteger">5</hasPriority><br>    </Person><br></owl:distinctMembers> |

**Figure 4. RDF/OWL format of Environment, Activity, Preference and Person data**

In addition to the environmental data, we also store the information of home members, and their preferences, activities as well. Each ontology class has the relationship in order that we can query information from these by using one OWL query language. The size of average dataset collecting daily is reduced remarkable. From Table II, the amount of information after excluding the relevant information as the threshold described above is 70 times smaller than without doing the exclusion.

**TABLE II**

**THE SIZE OF AVERAGE DATASET COLLECTING DAILY**

| Size (number of average KB for a day with 10 sensors) | |
|---|---|
| Without excluding the duplicated information | 115200 KB |
| Excluding the duplicated information | 1500 KB |

Applying query language is one of the most important steps in using ontologies. After building the dataset, the context aware system operates some queries from these created ontologies. In fact, we want to obtain the target is what kind of activity is suitable for the current context. In discussing context awareness, it is important to define context. A context aware reasoning system is able to answer these questions as sequence: who, what, when, where, and why. 'You' in this context assumes 'the user'.

- Who are you, and who are you with?
- Where are you?
- What do you intend to do?
- When do you do that?
- Why do you do that?

In order to evaluate rules and queries on context defined in the preceding part, we use the Jena reasoning engine. In addition, we choose SPARQL as the query language for accessing RDF/OWL, which is designed by the W3C RDF Data Access Working Group [17]. Figure 5 illustrates the common query and corresponding result in smart home, *"What are the statuses of home appliances satisfying the current users' preferences, for example, currently 6 pm?"*. Priority and time attributes play the main role in sorting the result set. In its turn, the ontology reasoning engine sends commands to home appliances controller.

| Query |
|---|
| SELECT DISTINCT ?person ?what ?appliance ?status ?priority |
| WHERE |
| { |
|     ?work :When :_180000. |
|     ?environment :hasTime :_180000. |
|     ?environment :personIn ?person. |
|     ?work :Who ?person. |
|     ?work :Do ?what. |
|     ?person :hasPriority ?priority. |
|     ?what ?appliance ?status. |
|     filter(datatype(?status)=xsd:boolean) |
| } |
| ORDER BY DESC(?priority) |

**Result**

| person | what | appliance | status | priority |
|---|---|---|---|---|
| Son | Self-study | TV | False | 5 |
| Son | Self-study | Air conditioner | True | 5 |
| Son | Self-study | Light | True | 5 |
| Son | Self-study | Projector | True | 5 |

**Figure 5. Query and reasoning result**

## 5. Conclusion and future works

In this paper, we introduce the context ontology implementation for smart home, the prototype of ubiquitous home network. We analyze the ability applying ontologies in the context aware system. Then we applied them in a real middleware framework, UTOPIA. We listed the facing issues and also the solutions to help us overcome them. This paper describes both what to do and how to do that. Our research is a step towards the standardization of a shared ontology for ubiquitous computing applications.

We believe by defining a shared ubiquitous computing ontology, we can help system developers to reduce their efforts in creating ontologies and to be more focused on the actual system implementations. The realization of context aware system for smart home can be improved with other different environments integration.

## 6. Acknowledgment

This research was supported by the Industry Promotion Project for Regional Innovation. The authors also would like to thank the BK21 Project of Korea and the anonymous reviewers for useful comments.